\documentclass[%
 reprint,
 amsmath,amssymb,
 aps,
 pra,
]{revtex4-2}

\usepackage{graphicx}
\usepackage{dcolumn}
\usepackage{bm}
\usepackage[hidelinks]{hyperref}

\usepackage{todonotes}
\usepackage{xcolor}
\usepackage{braket}
\usepackage{subcaption}
\usepackage{amsthm}

\begin{document}

\preprint{APS/123-QED}

\title{Generic and Scalable Differential Equation Solver for Quantum Scientific Computing}

\author{Jinhwan Sul}
\author{Yan Wang}%
 \email{yan-wang@gatech.edu}
\affiliation{%
 Georgia Institute of Technology,  Atlanta, Georgia 30332, USA}%


\date{\today}

\begin{abstract}
One of the most important topics in quantum scientific computing is solving differential equations. In this paper, generalized quantum functional expansion (QFE) framework is proposed. In the QFE framework, a functional expansion of solution is encoded into a quantum state and the time evolution of the quantum state is solved with variational quantum simulation (VQS).
The quantum functional encoding supports different numerical schemes of functional expansions.
The lower bound of the required number of qubits is double logarithm of the inverse error bound in the QFE framework.
Furthermore, a new parallel Pauli operation strategy is proposed to significantly improve the scalability of VQS. The number of circuits in VQS is exponentially reduced to only the quadratic order of the number of ansatz parameters.
Four example differential equations are solved to demonstrate the generic QFE framework.

\end{abstract}

\maketitle


\section{Introduction \label{sec:intro}}

Quantum scientific computing recently emerged as a new paradigm to solve difficult engineering and scientific problems based on quantum computers, including engineering simulation \cite{wang2013simulating, wang2016accelerating}, optimization \cite{ wang2014global, wang2023opportunities, kim2023quantum}, and materials design \cite{bauer2020quantum}. One of the most important topics in quantum scientific computing is solving differential equations. 
The main challenge, however, is that quantum computation is limited to unitary operations, whereas most of the differential equations describe non-unitary dynamics.
In order to fill this gap, the problem of solving differential equations is converted to other problems, such as solving linear equations~\cite{clader2013preconditioned,berry2014high,berry2017quantum,childs2020quantum,liu2021variational,sato2021variational,gnanasekaran2023efficient}, Hamiltonian simulation \cite{leyton2008quantum,engel2019quantum,costa2019quantum,gonzalez2023efficient,jin2023quantum,sato2024hamiltonian,brearley2024quantum}, and performing variational quantum simulation~\cite{fontanela2021quantum,kubo2021variational,alghassi2022variational,leong2023variational}.

However, the existing approaches face the issues of generalization and scalability. Hamiltonian simulation can only be applied to those problems that can be converted to the Schr\"{o}dinger equation. The conversion process is very problem-specific and difficult to be generalized. To utilize quantum linear equation solvers~\cite{harrow2009quantum,bravo2023variational}, differential equations should be discretized over space and time. The linearization of highly nonlinear regions requires a very large number of grid points, and thus does not scale well.
In existing approaches that are based on variational quantum simulation (VQS)~\cite{yuan2019theory, mcardle2019variational, endo2020variational}, only ordinary differential equations (ODEs) are solved.
The finite difference method is applied to discretize the spatial domain and convert partial differential equations (PDEs) to ODEs. Similarly, the linearization based on the finite difference method has the scalability issue.
It is desirable to develop generic and scalable approaches to solve a wide range of differential equations with good accuracy using quantum computers. 

Recently, a quantum functional expansion (QFE) framework~\cite{sul2024qfe} was proposed to solve differential equations based on VQS. 
In the QFE framework, functional expansion is first performed in the solution space, and the differential equation is converted to ODEs of coefficients.
The coefficients are encoded as the quantum states and obtained with VQS.
In this paper, we generalize the QFE framework 
by extending the functional expansion in both spatial domain and probability space to support different numerical schemes for solving deterministic and stochastic differential equations.
Furthermore, a new parallel Pauli operation strategy is proposed to significantly improve the scalability of VQS. As a result, the number of circuits in VQS is reduced from $\mathcal{O}(M^2 + 4^n M)$ to $\mathcal{O}(M^2)$, where $n$ is the number of qubits and $M$ is the number of ansatz parameters.

In VQS, running multiple quantum circuits is the major computational overhead.
The number of required quantum circuits is $(M+1)^2+P(M+1)$, which depends on the number of ansatz parameters $M$ and the number of unitaries $P$ after the Hamiltonian decomposition.
Therefore, it is important to decompose the Hamiltonian into $\mathcal{O}(\mathrm{poly}(n))$ number of unitaries, as in the originally proposed VQS~\cite{endo2020variational}.
Fontanela et al.~\cite{fontanela2021quantum} utilized VQS to solve PDEs. However, the Pauli decomposition method was used to decompose the Hamiltonian. In the worst-case scenario, a total number of $P=4^n$ unitaries are required as a result of the Pauli decomposition.
Kubo et al.~\cite{kubo2021variational} used VQS to solve 
stochastic differential equations. The trinomial tree approximation method was applied to discretize the spatial domain so that the differential equations are converted to systems of ODEs with sparse Hamiltonians.
A shift operator unitary decomposition approach was proposed to decompose the Hamiltonians into $\mathcal{O}(\mathrm{poly}(n))$ number of unitaries.
Alghassi et al.~\cite{alghassi2022variational} and Leong et al.~\cite{leong2023variational} utilized finite difference method to discretize the spatial domain, and the PDEs are converted to systems of ODEs. 
The finite difference method leads to sparse Hamiltonians, which are decomposed with the shift operator approach similarly.
Both the trinomial tree approximation and finite difference method can result in sparse Hamiltonians, and $(M+1)^2+\mathrm{poly}(n)(M+1)$ number of circuits are needed in VQS.
However, the accuracy of the solution is sensitive to the discretization of spatial domain. A large number of qubits are required to reduce the discretization error.
More importantly, reducing the number of circuits is critical to improve the scalability and applicability of VQS.

In this paper, we propose a parallel Pauli operation strategy to realize the linear combination of Pauli gates for VQS.
The required number of circuits is significantly reduced to $\mathcal{O}(M^2)$ regardless of the sparsity of Hamiltonians.
The linear combination of Pauli gates allows us to perform different Pauli operations with one single circuit.
The tensor products of Pauli matrices with real and imaginary coefficients are implemented separately.

The remainder of the paper is organized as follows.
The generalized framework of QFE is introduced in Section~\ref{sec:qfe}. 
The proposed parallel Pauli operation strategy for VQS is described in Section~\ref{sec:para}. 
Four different differential equations are solved in Section~\ref{sec:numerical} to demonstrate the generic QFE framework, including a stochastic differential equation, a deterministic heat equation, and a stochastic heat equation. In addition, a strongly coupled system of ODEs with dense Hamiltonian is used to demonstrate the scalability of the parallel Pauli operation strategy.
In Section~\ref{sec:discussion}, the state preparation and the error bound of QFE framework are discussed.
\section{Quantum Functional Expansion \label{sec:qfe}}

\begin{figure}
    \centering
    \includegraphics[width=\linewidth]{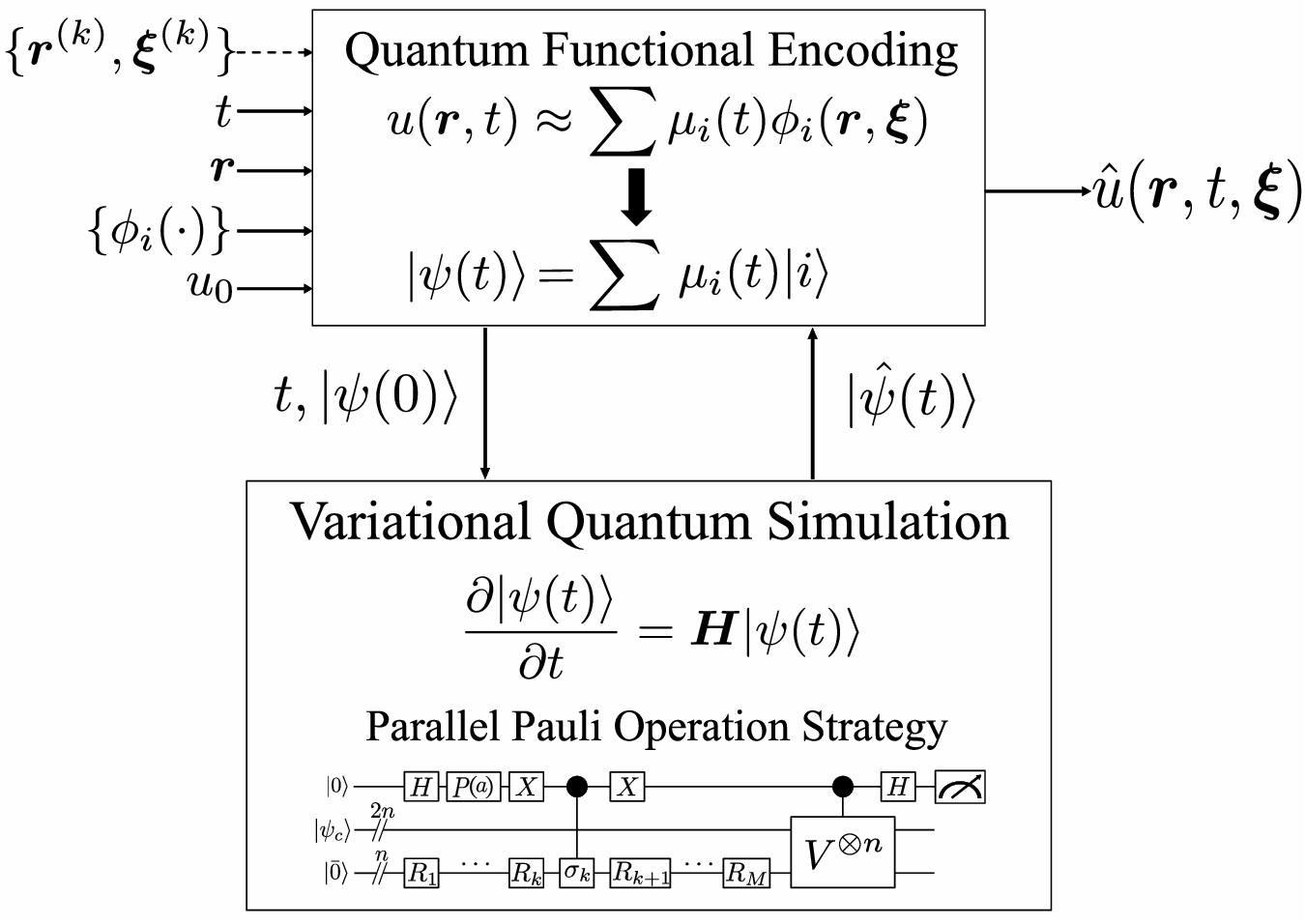}
    \caption{The proposed generic QFE framework.}
    \label{fig:QFEframework}  
\end{figure}

The QFE framework, as illustrated in Fig.~\ref{fig:QFEframework}, consists of two components.
The first component is quantum functional encoding, and the second component is VQS with the parallel Pauli operation strategy.
In quantum functional encoding, the solution of differential equations $u(\bm{r}, t)$ is approximated as a linear combination of basis functions. 
The functional expansion of the solution to stochastic or deterministic differential equations can be generally expressed as
\begin{equation}
    u(\bm{r}, t; \omega) \approx \sum_{i} \mu_i(t)\phi_i(\bm{r}, \bm{\xi}(\omega)),
\label{eq:functional_expansion}
\end{equation}
where $\phi_i(\cdot)$ is the $i$-th basis function, and $\mu_i$ is the corresponding expansion coefficient.
The functional basis $\{ \phi_i \}$ consists of tensor products of functional bases.
$\bm{r}$ is the spatial coordinates, and $\bm{\xi}$ is a vector of independent random variables.
For stochastic differential equations, the solution $u(\bm{r}, t; \omega)$ is also stochastic, where $\omega \in \Omega$ is a sample from a complete probability space $(\Omega, \mathcal{F}, P)$. Here, $\Omega$ is the set of events, $\mathcal{F}\subset 2^{\Omega}$ is the $\sigma$-algebra of events, and $P$ is a probability measure.
After the functional expansion in Eq.~(\ref{eq:functional_expansion}) is performed, the coefficients are encoded into a quantum state as
\begin{equation}
    \ket{\psi(t)} = \sum_i \mu_i(t) \ket{i},
\label{eq:quantum_functional_encoding}
\end{equation}
where $\ket{i}$'s are the basis states. 
The differential equation is converted to the system of ODEs of coefficients.
The second component of the QFE framework, VQS, is used to solve the system of ODEs.
In the generalized QFE framework, parallel Pauli operation strategy is utilized in VQS, which is discussed in Section~\ref{sec:para}.

The generalized QFE framework supports different numerical schemes, such as Galerkin and collocation methods, both deterministic and stochastic.
The functional expansions of these four numerical methods are special cases of Eq.~(\ref{eq:functional_expansion}).
When the Galerkin method is applied to solve differential equation, the solution is approximated as 
\begin{equation}
     u(\bm{r}, t) \approx \sum_{i} \mu_{i}(t)\phi_i(\bm{r}).
\label{eq:galerkin}
\end{equation}
In the stochastic Galerkin methods~\cite{ghanem2003stochastic, babuska2004galerkin}, 
random variables can be approximated with polynomial chaos expansion~\cite{xiu2002wiener} or Karhunen-Lo\'{e}ve expansion~\cite{loeve1977elementary}.
The solution of stochastic partial differential equations is approximated as
\begin{equation}
    u(\bm{r}, t) \approx \sum_{i, j} \mu_{i,j}(t)\phi_i(\bm{r}, \bm{\xi}).
\label{eq:FE_pce}
\end{equation}
Here, the functional basis $\{\phi_i(\bm{r}, \bm{\xi})\} = \{\chi_i(\bm{r}) \} \otimes \{ \eta_j(\bm{\xi})\}$ is generated from the tensor product of bases for deterministic spatial domain $\{\chi_i(\bm{r}) \}$ and probability space $\{ \eta_j(\bm{\xi})\}$.

When the collocation method~\cite{canuto2006spectral} is applied to solve deterministic differential equations, the functional expansion of the solution is
\begin{equation}
    u(\bm{r}, t) \approx \sum_{i} u(\bm{r}^{(i)},t)\phi_i(\bm{r}),
\label{eq:deter_col}
\end{equation}
where $\bm{r}^{(i)}$'s are the collocation points chosen in the spatial domain, and $\phi_i(\cdot)$'s are the interpolation functions that satisfy $\phi_i(\bm{r}^{(j)}) = \delta_{ij}$. 
In the collocation method, the interpolation functions are the basis, where the number of expansion terms is the number of collocation points.

For stochastic partial differential equations, the stochastic collocation method~\cite{babuvska2007stochastic} can be applied.
Stochastic collocation points $\bm{\xi}^{(i)}$'s are samples selected in the probability space.
Quadrature points are often selected as stochastic collocation points to compute the moments as  weighted summations.
As a result, a deterministic partial differential equation at each stochastic collocation point is solved.
The solution of the deterministic differential equation is approximated as
\begin{equation}
     u^{(i)}(\bm{r}, t) \approx \sum_{k} \mu_k(\bm{\xi}^{(i)}, t)\phi_k(\bm{r}), \,\, \forall \, i.
\end{equation}

For stochastic differential equations, we also propose to compute the mean and variance of the solution using the quantum amplitude estimation.
This is enabled by the quantum state encoding scheme in Eq.~(\ref{eq:quantum_functional_encoding}).
This is very efficient because quantum tomography is not needed. 
The solution of the stochastic differential equations is approximated as
\begin{equation}
    u(t) = \sum_{i=0}^{Q-1} \mu_i(t) \phi_i(\xi),
\end{equation}
where the functional expansion is truncated with $Q$ terms, and $\phi_0(\cdot) = 1$. 
If the weight corresponding to the orthogonal functional basis $\{ \phi_i(\cdot)\}$ is 
the probability distribution function of $\xi$, the mean and variance of $u(t)$ can be obtained from the expansion coefficients from the orthogonality.
The first moment is computed as
\begin{equation}
    \mathbb{E}[u(t)] = \sum_{i=0}^{Q-1} \mu_i(t) \mathbb{E}[\phi_i(\xi)] = \mu_i(t) \langle \phi_0, \phi_i \rangle = \mu_0(t).
\label{eq:first_moment}
\end{equation}
The variance is obtained from the orthogonality as
\begin{align}
    \mathrm{Var}(u(t)) &= \mathbb{E}[ (\sum_{i=1}^{Q-1} \mu_i(t)h_i(\xi) )^2]  = \sum_{i,j=1}^{Q-1} \mu_i(t)\mu_j(t) \langle \phi_i, \phi_j \rangle \nonumber \\ &= \sum_{i=1}^{Q-1} \mu_i^2(t) = 1-\mu_0^2(t).
\label{eq:second_moment}
\end{align}
$\mu_0(t)$ is the probability amplitude of $\ket{\bar{0}}$ in Eq.~(\ref{eq:quantum_functional_encoding}). As a result, the first and second moments can be obtained from the quantum amplitude estimation~\cite{brassard2002quantum}. Given the desired error bound $\varepsilon$, $\mathcal{O}(1/\varepsilon)$ number of shots are required, which is a quadratic speed up over quantum tomography.
\section{Parallel Pauli Operation Strategy for Variational Quantum Simulation \label{sec:para}}

VQS can be used to solve the general time evolution problem 
\begin{equation}
    \frac{\partial \ket{\psi(t)}}{\partial t} = \mathbf{H} \ket{\psi(t)},
\end{equation}
where the Hamiltonian $\mathbf{H}$ is not necessarily a Hermitian operator.
The unnormalized quantum state $\ket{\psi(t)}$ is estimated as $\ket{\hat{\psi}(\bm{\theta}(t))} = \alpha(t) U(\bm{\beta}(t)) \ket{0}^{\otimes n}$ with a parameterized ansatz $U(\cdot)$ and a normalization factor $\alpha$. The normalization factor and ansatz parameters are redefined as a single parameter vector $\bm{\theta}(t) = (\alpha(t), \bm{\beta}(t)) = (\theta_0(t), \theta_1(t), \dots, \theta_M(t))$.
By using the ansatz, the generalized time-evolution is mapped to the time-evolution of the ansatz parameters. 
The ansatz parameters are obtained with the McLachlan's variational principle~\cite{mclachlan1964variational}
\begin{equation}
    \delta \|  \frac{\partial}{\partial t}\ket{\hat{\psi}(\bm{\theta}(t))} - \mathbf{H} \ket{\hat{\psi}(\bm{\theta}(t))} \| = 0,
\label{eq:mclachlan}
\end{equation}
where $\| \ket{\pi} \| = \sqrt{\langle \pi | \pi \rangle}$ for any quantum state $\ket{\pi}$. 
Eq.~(\ref{eq:mclachlan}) is equivalent to
\begin{equation}
    \sum_{i=0}^M A_{ik}(t) \dot{\theta_i} = C_k(t), \,\, k=0,\,1, \ldots,\, M.
\label{eq:linear}
\end{equation}
where
\begin{equation}
    A_{ik}(t) = \Re \Bigl\{ \frac{\partial \bra{\hat{\psi}(\bm{\theta}(t))}}{\partial \theta_k} \frac{\partial \ket{\hat{\psi}(\bm{\theta}(t))}}{\partial \theta_i} \Bigr\},
\label{eq:A_ik}
\end{equation}
\begin{equation}
    C_k(t) = \Re \Bigl\{ \frac{\partial \bra{\hat{\psi}(\bm{\theta}(t))}}{\partial \theta_k} \mathbf{H} \ket{\hat{\psi}(\bm{\theta}(t))} \Bigr\}.
\label{eq:C_k}
\end{equation}
$A_{ik}$ and $C_k$ are obtained from multiple modified Hadamard test quantum circuits. 
A total $(M+1)^2$ number of quantum circuits are required to obtain $A_{ik}$ for all $i$ and $k$, where each $A_{ik}$ is calculated from one circuit.
In order to compute one $C_k$, the Hamiltonian is first decomposed into $P$ unitaries as $\mathbf{H} = \sum_{i=1}^{P} c_i U_{i}$.
$C_k$ is then computed from
\begin{equation}
    C_k(t) = \sum_{i=1}^{P} \Re \Bigl\{ c_i \frac{\partial \bra{\hat{\psi}(\bm{\theta}(t))}}{\partial \theta_k} U_{i} \ket{\hat{\psi}(\bm{\theta}(t))} \Bigr\}.
\label{eq:C_k_sum}
\end{equation}
The weighted sum of the $P$ terms in Eq.~(\ref{eq:C_k_sum}) is calculated on a classical computer. 
As a result, the total number of required quantum circuits to obtain all the $C_k$'s is $P(M+1)$.

In the original VQS~\cite{endo2020variational}, 
the required number of circuits scales as $\mathcal{O}(M^2 + 4^n M)$.
If the Hamiltonian is an arbitrary dense matrix, the Pauli decomposition is
\begin{equation}
    \mathbf{H} = \sum_{i=1}^{4^n} \gamma_i \sigma_i^{(n)},
\label{eq:pauli_decomposition}
\end{equation}
where $\sigma_i^{(n)}$'s are the $n$-tensor products of Pauli matrices and $\gamma_i$'s are complex coefficients.
In the worst-case scenario with a dense Hamiltonian, the number of non-zero terms in the Pauli decomposition can be $4^n$.
As a result, the total required number of circuits increases exponentially. Thus, the scalability of the original VQS is the major barrier for its feasibility of solving real-world problems.

\begin{figure}
    \centering    
    \includegraphics[width=\linewidth]{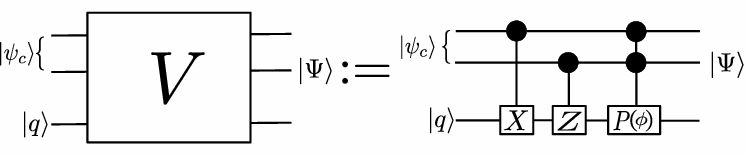}
    \caption{Quantum circuit to compute linear combination of single qubit Pauli gates. $\phi = -\pi/2$.}
    \label{fig:para_qfe_building_block}
\end{figure}

The proposed parallel Pauli operation strategy significantly improves the scalability of VQS. 
The new strategy is to compute the linear combination of Pauli operations with a single circuit based on quantum parallelism.
The linear combination of single-qubit Pauli gates can be implemented with the quantum circuit in Fig.~\ref{fig:para_qfe_building_block}. 
This circuit, denoted as $V$, consists of three controlled unitary operations, starting with a controlled $X$ gate, followed by a controlled $Z$ gate, and a controlled phase shift gate with a phase shift $-\pi/2$.
Different Pauli operations are applied to the target qubit $\ket{q}$ depending on the basis state of the ancillary controlled qubits $\ket{\psi_c}$. 
If the quantum state of the ancillary qubits is prepared as
\begin{equation}
    \ket{\psi_c} = c_1\ket{00} + c_2\ket{01} + c_3\ket{10} + c_4\ket{11},
\end{equation}
then the output quantum state is 
\begin{align}
\begin{split}
    \ket{\Psi} &= \ket{00} \otimes c_1 I\ket{q} + \ket{01} \otimes c_2 Z\ket{q}  \\  &+ \ket{10} \otimes c_3 X\ket{q} + \ket{11} \otimes c_4 Y\ket{q}.
\end{split}
\end{align}
For $n$-qubit Pauli operations, $V^{\otimes n}$ is applied.

\begin{figure}
    \captionsetup[subfigure]{justification=justified}
    \begin{subfigure}{\linewidth}    
        \includegraphics[width=\linewidth]{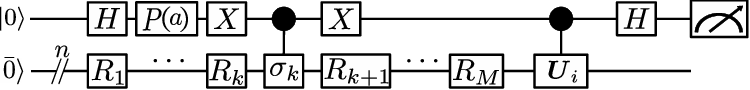}
        \caption{}
        \label{fig:NoPara_MHC}
    \end{subfigure}
    \par
    \bigskip
    \begin{subfigure}{\linewidth}
        \centering    
        \includegraphics[width=\linewidth]{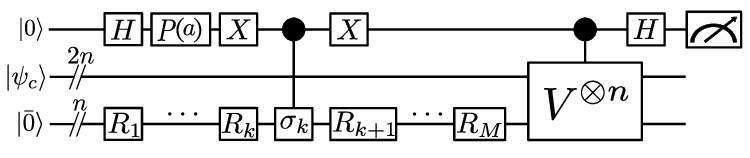}
        \caption{}
        \label{fig:Para_MHC}
    \end{subfigure}
    \caption{Modified Hadamard test circuits for VQS. (a) Quantum circuit in the original VQS to compute one of the terms in Eq.~(\ref{eq:C_k_sum}). (b) Quantum circuit of the new proposed parallel Pauli operation strategy to compute $C'_k(t)$.}
\label{fig:para_whole}
\end{figure}

The comparison between the scalable VQS with the parallel Pauli operation strategy and the original VQS is shown in Fig.~\ref{fig:para_whole}.
Fig.~\ref{fig:NoPara_MHC} is the circuit used in the original VQS, and Fig.~\ref{fig:Para_MHC} is the circuit with the new parallel Pauli operation strategy. 
Here, $R_k$ is a Pauli rotation gate with a rotation angle of $\theta_k$, and $\sigma_k$ is the Pauli gate corresponding to the rotation axis of $R_k$.
The difference between quantum circuits in Fig.~\ref{fig:NoPara_MHC} and Fig.~\ref{fig:Para_MHC} is that the unitary $U_i$ is substituted with $V^{\otimes n}$ in the parallel Pauli operation strategy quantum circuit.
In addition, $2n$ ancillary qubits are introduced to create the superposition of $n$-qubit Pauli operations.
The quantum state of ancillary qubits is
\begin{equation}
    \ket{\psi_c} = \sum_{i=1}^{2^{2n}} \sqrt{c_i}\ket{i},
\label{eq:ppo_sp}
\end{equation}
which is used to compute 
\begin{equation}
    C'_k(t) = e^{i a} \Re \Bigl\{ \frac{\partial \bra{\hat{\psi}(\bm{\theta}(t))}}{\partial \theta_k} \sum_{i=1}^{2^{2n}} c_i \sigma_i^{(n)} \ket{\hat{\psi}(\bm{\theta}(t))} \Bigr\}.
\end{equation}
In order to compute $C'_k(t)$ correctly, the coefficients $c_i$'s need to have positive real values and satisfy the normalization condition
\begin{equation}
    \sum_{i=1}^{2^{2n}} c_i = 1.
\label{eq:ck_condition_norm}
\end{equation}

Since the coefficients $\gamma_i$'s in Eq.~(\ref{eq:pauli_decomposition}) have complex values, the Hamiltonian needs to be decomposed to
\begin{equation}
    \mathbf{H} = g_1 \mathbf{H}_1 - g_2 \mathbf{H}_2 +  g_3 \sqrt{-1} \mathbf{H}_3 - g_4 \sqrt{-1} \mathbf{H}_4,
\end{equation}
where $\mathbf{H}_l$'s are the linear combinations of Pauli operations with positive coefficients. 
The real-valued scale factors $g_l$'s are positive in order to ensure $\mathbf{H}_l$'s to satisfy the normalization condition in Eq.~(\ref{eq:ck_condition_norm}).
As a result of the decomposition, the real and imaginary parts of the complex coefficients $\gamma_i$'s, with positive and negative values, are separately computed.
Therefore, $C_k(t)$ is obtained as
\begin{align}
\begin{split}
    C_k(t) &= g_1 \Re \Bigl\{ \frac{\partial \bra{\hat{\psi}(\bm{\theta}(t))}}{\partial \theta_k} \mathbf{H}_1 \ket{\hat{\psi}(\bm{\theta}(t))} \Bigr\} \\
    &- g_2 \Re \Bigl\{ \frac{\partial \bra{\hat{\psi}(\bm{\theta}(t))}}{\partial \theta_k} \mathbf{H}_2 \ket{\hat{\psi}(\bm{\theta}(t))} \Bigr\} \\
    &+ g_3 \Bigl( \sqrt{-1}\Im \Bigl\{ \frac{\partial \bra{\hat{\psi}(\bm{\theta}(t))}}{\partial \theta_k} \mathbf{H}_3 \ket{\hat{\psi}(\bm{\theta}(t))} \Bigr\} \Bigr) \\
    &- g_4 \Bigl( \sqrt{-1}\Im \Bigl\{ \frac{\partial \bra{\hat{\psi}(\bm{\theta}(t))}}{\partial \theta_k} \mathbf{H}_4 \ket{\hat{\psi}(\bm{\theta}(t))} \Bigr\} \Bigr).
\end{split}
\end{align}
A total $4M$ number of circuits are required to compute all the $C_k$'s. Hence, only $\mathcal{O}(M^2)$ number of quantum circuits are required in VQS with the proposed parallel Pauli operation strategy for any Hamiltonian.
\section{ Demonstrations \label{sec:numerical}}

Four differential equations are solved to demonstrate the QFE framework and the parallel Pauli operation strategy. All the quantum simulations are performed with a state vector simulator excluding the effect of quantum noise. 

\subsection{Strongly Coupled System of Ordinary Differential Equations \label{subsec:dense_ode}}

\begin{figure}
    \centering    
    \includegraphics[width=\linewidth]{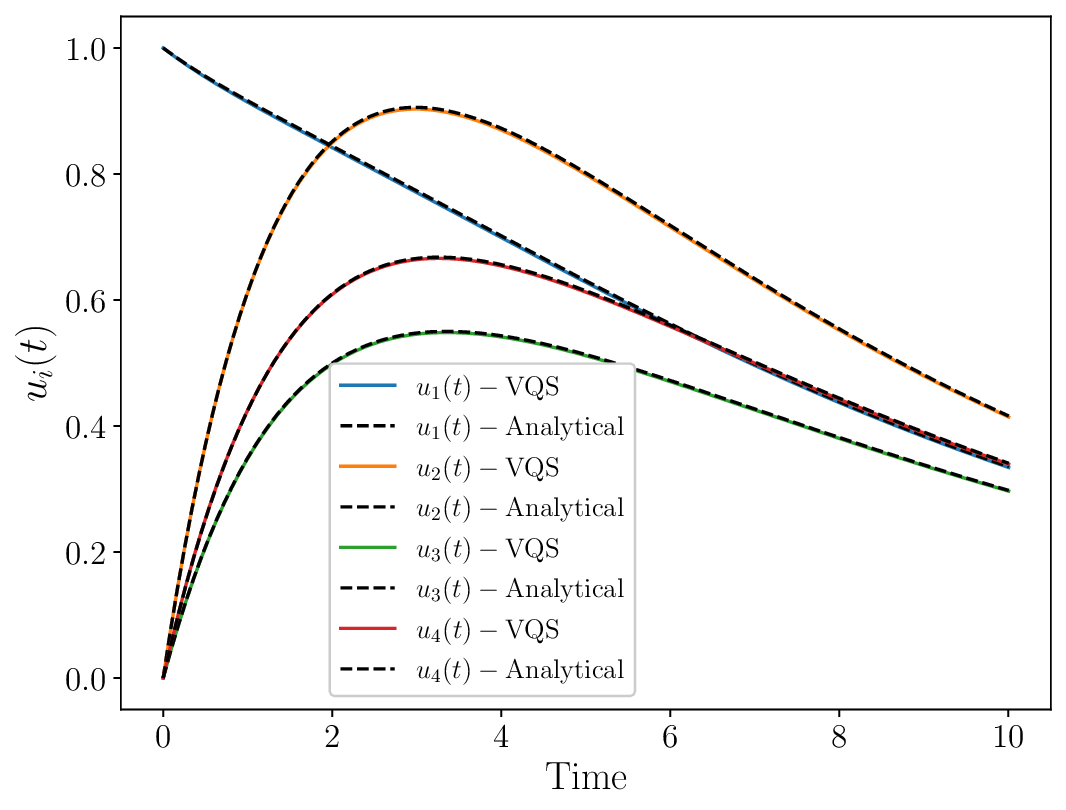}
    \caption{The solutions of ODEs with the dense system matrix.}
    \label{fig:dense_ode}
\end{figure}

A strongly coupled system of ordinary differential equations is solved to demonstrate the scalable VQS with the parallel Pauli operation strategy. 
The system of ODEs is
\begin{equation}
    \frac{d \bm{u}(t)}{dt} = \mathbf{A}\bm{u}(t),
\label{eq:system_of_ODEs}
\end{equation}
where $\bm{u}(t) = [u_1(t), u_2(t), u_3(t), u_4(t)]^\intercal$, and the initial condition is $\bm{u}(0) = [1,0,0,0]^\intercal$. $n=2$ qubits are used to encode the solution $\bm{u}(t)$.
The system matrix
\begin{equation}
    \mathbf{A} = \begin{bmatrix}
    -0.1 & 0.4 & 0.2 & -0.7 \\
    0.9 & 0.1 & -0.1 & -1.1 \\
    0.5 & 0.2 & -0.4 & -0.5 \\
    0.6 & 0.5 & 0.3 & -1.6 \\
    \end{bmatrix}
\end{equation}
is a dense matrix. Therefore, $u_i$'s are strongly correlated to each other.
The system matrix $\mathbf{A}$ is not necessarily Hermitian nor unitary.
Eq.~(\ref{eq:system_of_ODEs}) is difficult to solve using the original VQS, since an efficient unitary decomposition of $\mathbf{A}$ is absent. Moreover, the Pauli decomposition of $\mathbf{A}$ includes $4^n$ tensor products of Pauli matrices with non-zero coefficients.
In this example, an ansatz with $M=4$ parameters is used.
In order to solve Eq.~(\ref{eq:system_of_ODEs}) with the original VQS, $(M+1)^2 + 4^n (M+1) = 105$ quantum circuits are required. 
However, for the scalable VQS with the parallel Pauli operation strategy, only $(M+1)^2 + 4(M+1) = 45$ quantum circuits are needed. 
Four ancillary qubits are included to encode the Pauli decomposition coefficients. 
The results are shown in Fig.~\ref{fig:dense_ode}. It is seen that the scalable VQS can provide very accurate solutions in comparison with the analytical ones.

\subsection{Stochastic Ordinary Differential Equation \label{subsec:sode}}
The stochastic ODE is solved with the QFE framework using the stochastic Galerkin method.
The stochastic ODE is
\begin{equation}
    \frac{d u(t;\omega)}{dt} = a(\omega) u(t; \omega),
\label{eq:stochastic_ODE}
\end{equation}
where the initial condition is $u(0) = 1$, and $a$ is a Gaussian random variable $a \sim \mathcal{N}(-1.8, 1)$. 
The analytical solution of the mean and variance are 
\begin{align}
    \mathbb{E}[u(t)] &= u_0 e^{-1.8 t + \frac{t^2}{2}}, \\
    \mathrm{Var}[u(t)] &= u_0^2 e^{-3.6 t + t^2} (e^{t^2} - 1),
\end{align}
respectively.
Since the random variables are normally distributed, Hermite polynomials are chosen as the functional basis.

The polynomial chaos expansion is applied to $u(t)$ and $a$ as 
\begin{align}
    a &\approx \sum_{i=0}^{N-1} a_i h_i(\xi), \label{eq:sode_a_pce} \\
    u(t) &\approx \sum_{i=0}^{N-1} \mu_i(t) h_i(\xi),
    \label{eq:sode_u_pce}
\end{align}
where $h_{i}$ is the $i$-th order Hermite polynomial, and $\xi$ is the standard normal random variable.
Eq.~(\ref{eq:stochastic_ODE}) is transformed into
\begin{equation}
    \frac{d \mu_l(t)}{dt} = \sum_{i=0}^{N-1} \sum_{j=0}^{N-1} a_i \mu_j \frac{\langle h_l, h_i h_j \rangle}{\langle h_l, h_l \rangle},
\label{eq:sode_galerkin}
\end{equation}
for $l=0,\,1,\, \ldots,\, N-1$. The inner product between two functions $f(\cdot)$ and $g(\cdot)$ is defined as 
\begin{equation}
    \langle f, g \rangle = \int_{\mathcal{D}} f(\zeta) g(\zeta) w(\zeta) d\zeta,
\end{equation}
where $w(\cdot)$ is a weight function and $\mathcal{D}$ is the support of the functions.
Eq.~(\ref{eq:sode_galerkin}) forms the system of ODEs to be solved with the VQS.

\begin{figure}
    \captionsetup[subfigure]{justification=justified}
    \begin{subfigure}{\linewidth}    
        \includegraphics[width=\linewidth]{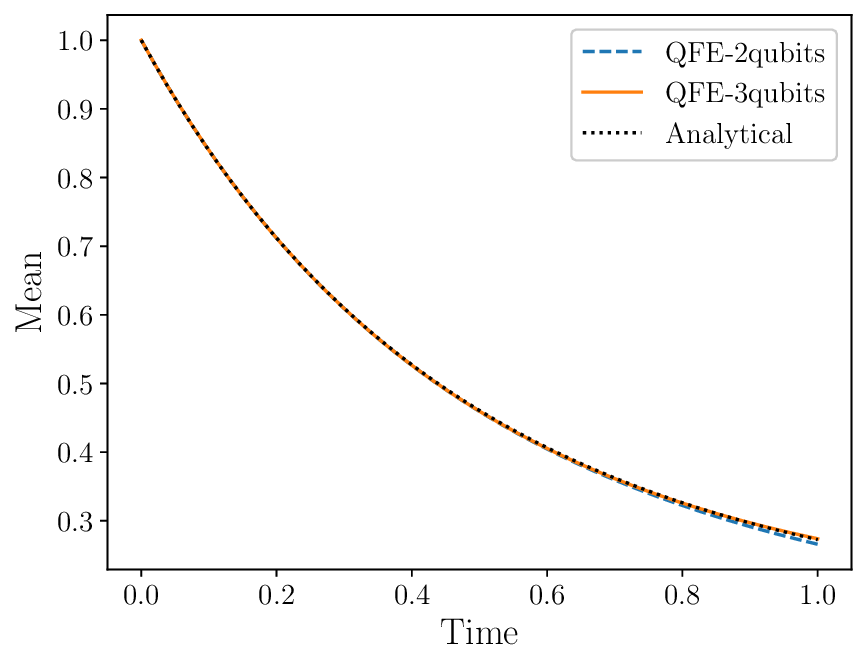}
        \caption{}
        \label{fig:sode_mean}
    \end{subfigure}
    \par
    \bigskip
    \begin{subfigure}{\linewidth}
        \centering    
        \includegraphics[width=\linewidth]{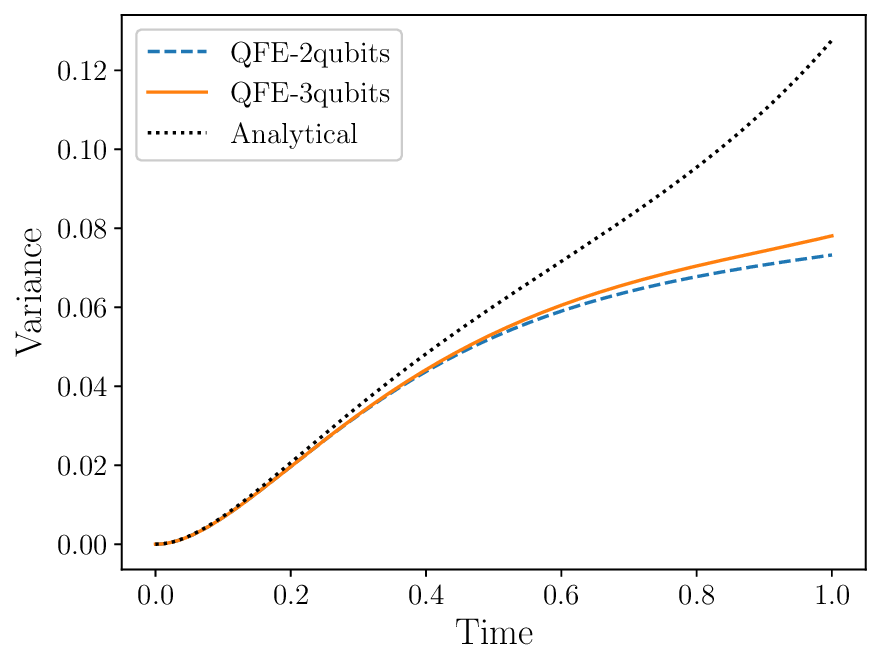}
        \caption{}
        \label{fig:sode_var}
    \end{subfigure}
    \caption{Comparison between the solutions obtained with the QFE framework and the analytical ones for the stochastic ODE.    
    (a) Mean of $u(t)$. (b) Variance of $u(t)$.}
\label{fig:sode_result}
\end{figure}

The mean and variance of the solution are obtained efficiently with quantum amplitude estimation from Eq.~(\ref{eq:first_moment}) and Eq.~(\ref{eq:second_moment}), respectively.
The results are shown in Fig.~\ref{fig:sode_mean} and Fig.~\ref{fig:sode_var}. The means and variances obtained from the QFE framework using $n=2$ and $n=3$ qubits are compared with the analytical solution.
The means match very well with the analytical solution even with a small number of qubits. The variances calculated with the QFE framework are accurate in the initial stage of the simulation.  The deviation from the analytical solution occurs as the number of time steps increases. 
As the number of qubits increases, the QFE result approaches closer to the analytical solution. More qubits can result in more accurate solutions. Because the number of qubits corresponds to the number of terms in the functional expansion.

\subsection{Heat Equation \label{subsec:heat}}

The generalized QFE framework is applied to solve the heat equation
\begin{equation}
    \frac{\partial u}{\partial t} = \alpha \frac{\partial^2 u}{\partial x^2} ,\quad x \in [-1, 1],
\label{eq:heat_eq}
\end{equation}
The Dirichlet boundary conditions are $u(\pm1, t)=0$, and the initial condition is $u(x,0) = \sin(\pi x)$. The thermal diffusivity is $\alpha = 0.3$. The analytical solution is $u(x,t) = e^{-\alpha \pi^2 t} \sin (\pi x)$.
The Chebyshev spectral collocation method~\cite{tang2006spectral} is utilized in the QFE framework.
The Gauss-Chebyshev-Lobatto points, $x_j = \cos(\pi j / N)$ ($0\leq j \leq N$) are chosen as the collocation points.
The solution of the equation is interpolated between the collocation points as
\begin{equation}
    u(x,t) \approx \sum_{j=0}^{N} u_j(t) S_j(x),
\label{eq:chebyshev_collocation}
\end{equation}
where $u_j(t)$ is the solution at the $j$-th collocation point $x_j$, and $S_j(x)$ is the $j$-th interpolation polynomial.
Therefore, Eq.~(\ref{eq:heat_eq}) is converted to
\begin{equation}
    \frac{\partial u_j(t)}{\partial t} = \alpha \sum_{l=0}^N (\mathbf{D}_N^{(2)})_{jl} u_l(t), \,\, (j=0, 1, \ldots, N),
\label{eq:ODEs_heat_equation}
\end{equation}
with the second-derivative Chebyshev differentiation matrix $\mathbf{D}_N^{(2)}$. More detailed derivation of the Chebyshev differentiation matrix is explained in Appendix~\ref{appendix:collocation}.

\begin{figure}
    \centering    
    \includegraphics[width=\linewidth]{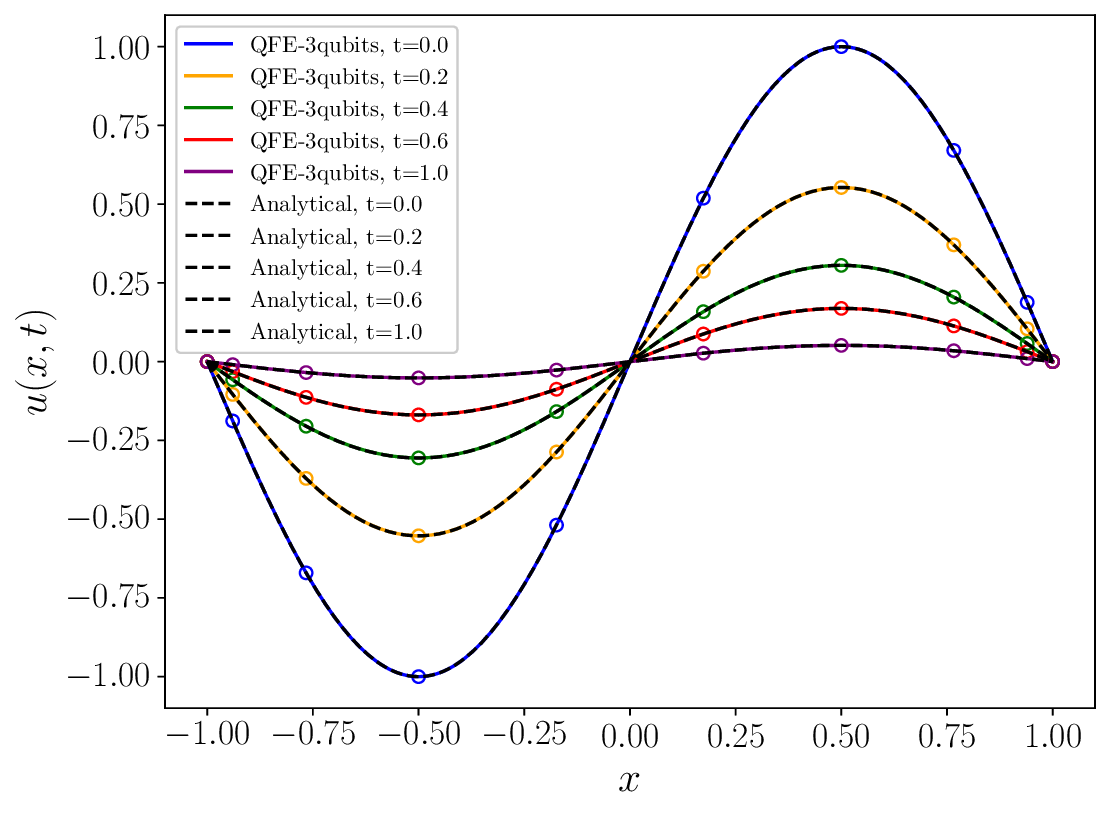}
    \caption{
    Comparison between the solutions obtained with the QFE framework and the analytical ones for the heat equation, where the dots indicate the solutions at the collocation points.}
    \label{fig:cheby_heat_col}
\end{figure}

The VQS with parallel Pauli operation is applied to solve Eq.~(\ref{eq:ODEs_heat_equation}). 
Since $\alpha\mathbf{D}_N^{(2)}$ is a dense matrix, the parallel Pauli operation strategy can significantly reduce the required number of circuits.
Here, a total number of $N=10$ collocation points are chosen, including two points at the boundaries. Eq.~(\ref{eq:ODEs_heat_equation}) is solved for the eight collocation points in the middle, since the solution at the boundaries are already determined from the boundary condition. A total of $n=3$ qubits are utilized. The result of the QFE framework is shown in Fig.~\ref{fig:cheby_heat_col}, where the solutions match the analytical ones well.

\subsection{Stochastic Heat Equation \label{subsec:stochastic_heat}}

The QFE framework is further demonstrated with the stochastic heat equation
\begin{equation}
    \frac{\partial u(x,t;\omega)}{\partial t} = \frac{\partial}{\partial x} \left( \gamma(x; \omega) \frac{\partial u(x,t;\omega)}{\partial x} \right),\, x \in [-1, 1].
\label{eq:stochastic_heat}
\end{equation}
The Dirichlet boundary conditions are $u(\pm1, t)=0$, and the initial condition is $u(x,0) = \sin(\pi x)$. The coefficient $\gamma(x)$ is a Gaussian random field with mean $\mu(x) = 2.7 - 0.1\sin(\pi x)$ and covariance function $C(x,x') = \exp(-\frac{1}{2}\|x - x'\|^2)$.

Karhunen-Lo\'{e}ve expansion is applied to approximate $\gamma(x)$ with the vector of independent standard normal random variables $\bm{\xi}$.
Stochastic collocation points $\{ \bm{\xi}^{(j)} \}$ are chosen from the probability space. The Gaussian-Hermite quadrature points are selected as the collocation points in this example.
For each collocation point $\bm{\xi}^{(j)}$, a deterministic heat equation
\begin{equation}
    \frac{\partial u^{(j)}(x,t)}{\partial t} = \frac{\partial}{\partial x} \left( \gamma^{(j)}(x) \frac{\partial u^{(j)}(x,t)}{\partial x} \right)
\label{eq:deterministic_heat}
\end{equation}
is solved, where $\gamma^{(j)}(x)$ is a realization of $\gamma(x;\omega)$ at the $j$-th collocation point.
Eq.~(\ref{eq:deterministic_heat}) is converted to a system of ODEs with the Chebyshev differentiation matrix. More detailed derivation is available in Appendix~\ref{appendix:scm}.

\begin{figure}
    \captionsetup[subfigure]{justification=justified}
    \begin{subfigure}{\linewidth}    
        \includegraphics[width=\linewidth]{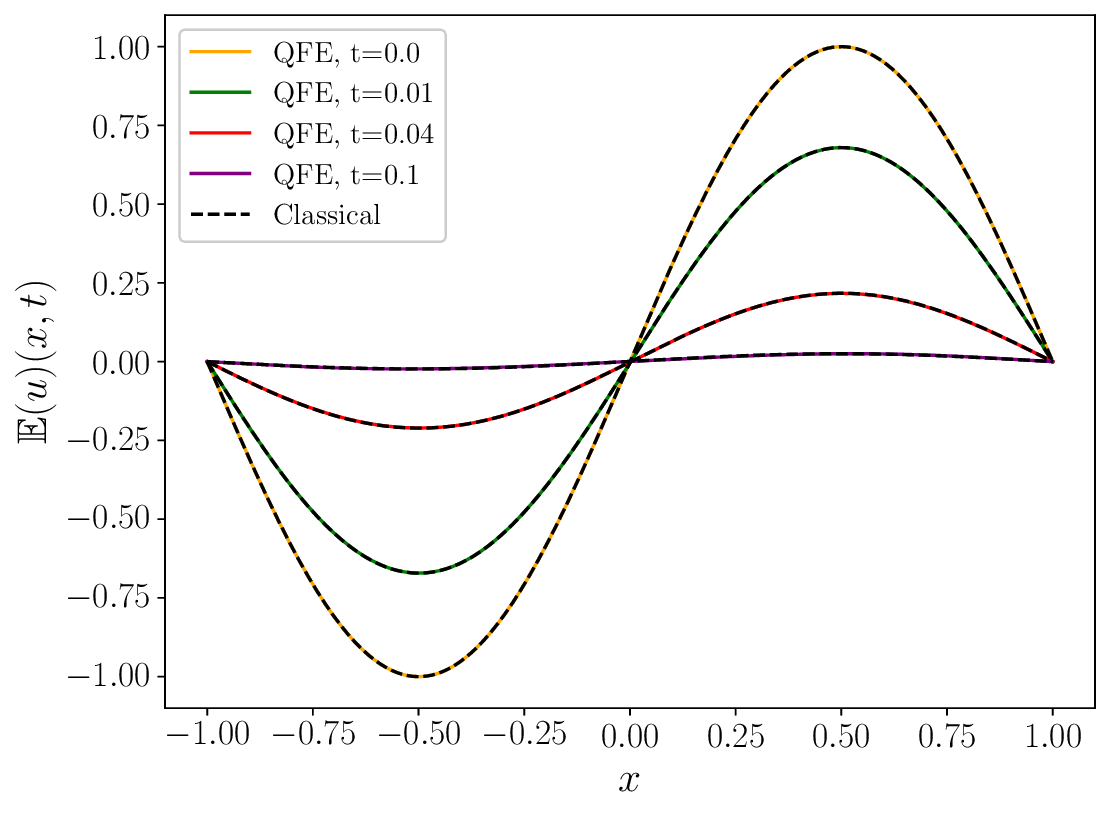}
        \caption{}
        \label{fig:scm_mean}
    \end{subfigure}
    \par
    \bigskip
    \begin{subfigure}{\linewidth}
        \centering    
        \includegraphics[width=\linewidth]{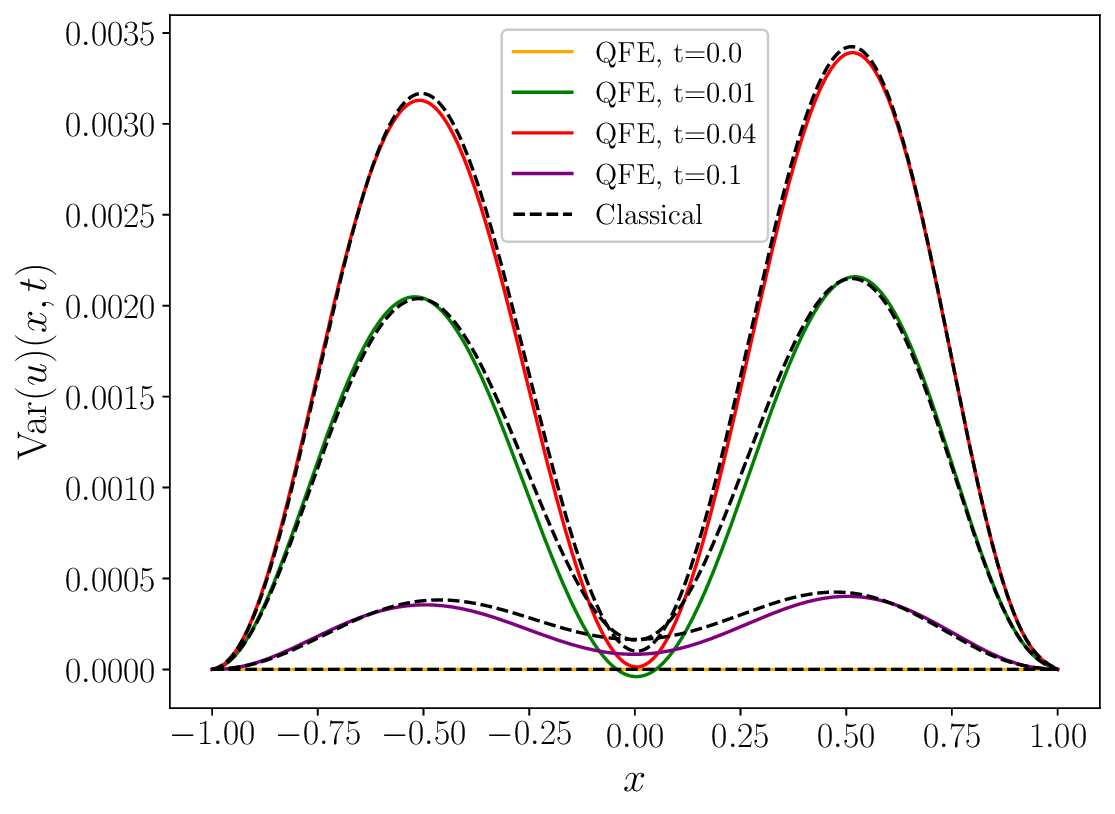}
        \caption{}
        \label{fig:scm_var}
    \end{subfigure}
    \caption{Comparison between the solutions obtained with the QFE framework and the analytical ones for the stochastic heat equation.
    (a) Mean of $u(x,t)$. (b) Variance of $u(x,t)$.}
\label{fig:scm_whole}
\end{figure}

The mean and variance of the solution are shown in Fig.~\ref{fig:scm_mean} and Fig.~\ref{fig:scm_var}, respectively.
The QFE result is compared with the solution obtained from classical computer based on the stochastic collocation method. 
The means of the two solutions match very well. 
Small deviations are observed for the variances near $x=0$.
This is mainly because the Gauss-Chebyshev-Lobatto collocation points are sparse near $x=0$ and dense at the boundaries.
\section{Discussions ~\label{sec:discussion}}

\subsection{Error Analysis and Required Number of Qubits ~\label{subsec:complexity_analysis}}

In the QFE framework, the number of qubits is directly related to the number of utilized basis functions and the accuracy of the solution. 
Thus, the lower bound of the required number of qubits can be obtained from the desired error bound $\varepsilon$.
An appropriate choice of functional basis can lead to the exponential reduction of truncation errors for smooth functions in the functional expansion.
As a result, the required number of qubits can be significantly reduced in the QFE framework.

The truncation error bound of a polynomial functional expansion in the Galerkin method is related to the smoothness of the function.
Let $P^{N}_{w^{\alpha,\beta}}$ be a polynomial functional expansion of $u(x) \in H^m_{w^{\alpha, \beta}}(\Omega)$ up to the $N$-th order, where
$H^m_{w^{\alpha, \beta}}(\Omega)$ denotes the $m$-differentiable weighted Sobolev space with the domain $\Omega = (-1, 1)$, and the weight function is $w^{\alpha, \beta}(x) = (1-x)^{\alpha}(1+x)^{\beta}$.
Then, for any $\alpha, \beta \in \mathbb{Z}$, the truncation error bound is~\cite{tang2006spectral}
\begin{equation}    
\| P^{N}_{w^{\alpha,\beta}} - u \|_{w^{\alpha,\beta}} \lesssim N^{-m} \|\partial_x^m u\|_{w^{\alpha+m,\beta+m}}.
\label{eq:truncation_bound_galerkin}
\end{equation}
Eq.~(\ref{eq:truncation_bound_galerkin}) indicates that the truncation error is $\mathcal{O}(N^{-m})$, and the error is dependent on the smoothness of the function. 
Since the required number of qubits is $n = \mathcal{O}(\log(N))$, the lower bound of the required number of qubits is $\mathcal{O}(1/m \log(\varepsilon^{-1}))$. For smooth functions with the exponential convergence, the lower bound of the number of qubits is $\mathcal{O}(\log \log(\varepsilon^{-1}))$.

The Jacobi polynomial interpolation error for the spectral collocation method is bounded similarly. 
Let the Jacobi polynomials $J_N^{\alpha, \beta}(x)$ be the interpolation functions, collocation points $\{x_j\}_{j=0}^N$ be the roots of $(1-x^2)\partial_x J_N^{\alpha, \beta}(x)$, and $I^{N}_{w^{\alpha,\beta}}$ be the interpolated estimation of $u(x) \in H^m_{w^{\alpha, \beta}}(\Omega)$ with respect to the collocation points $\{x_j\}_{j=0}^N$.
The polynomial interpolation error is ~\cite{tang2006spectral}
\begin{equation}    
\| I^{N}_{w^{\alpha,\beta}} - u \|_{w^{\alpha,\beta}} \lesssim N^{-m} \|\partial_x^m u\|_{w^{\alpha+m,\beta+m}}.
\label{eq:interpolation_error}
\end{equation}
Eq.~(\ref{eq:interpolation_error}) also indicates the error bound is $\mathcal{O}(N^{-m})$. Similarly, the lower bound of the required number of qubits is $\mathcal{O}(1/m \log(\varepsilon^{-1}))$, and for smooth functions, the required number of qubits is $\mathcal{O}(\log \log(\varepsilon^{-1})) \leq n$. 

For the stochastic differential equations, the truncation error of polynomial chaos expansion decreases exponentially when the weight function matches the probability distribution function~\cite{xiu2002wiener}.
Specifically, the truncation error bound of the polynomial chaos expansion for the $m$-differentiable $u \in L_2(\Omega, \mathcal{F}, P)$ with the Hermite polynomials is~\cite{augustin2008polynomial} 
\begin{equation}
    \left\| u - \sum_{i=0}^{N} \mu_i h_i \right\| \leq \frac{\| \partial_{x}^m u \|}{\prod_{i=0}^{m-1} \sqrt{N - i + 1}}.
\label{eq:hermite}
\end{equation}
The truncation error bound is $\mathcal{O}((N!)^{-1/2})$. Since $N! \sim \sqrt{2 \pi N}(N/e)^N$, the truncation error decreases exponentially. As a result, the lower bound of the required number of qubits is $\mathcal{O}(\log \log (\varepsilon^{-1}))$.

In summary, if the functional expansion of the solution exists and the solution is a smooth function, the lower bound of the required number of qubits is double logarithm of the inverse error bound in the QFE framework.

\subsection{State Preparation ~\label{subsec:state_preparation}}

In the proposed parallel Pauli operation strategy, the quantum state in Eq.~(\ref{eq:ppo_sp}) should be prepared.
The state preparation problem is defined as follows.
Given an $L_2$ normalized complex vector $\bm{\nu} = (\nu_0, \nu_1, \dots, \nu_{2^n - 1}) \in \mathbb{C}^{2^n}$ where $\sum_{i=0}^{2^n-1} |\nu_i|^2 = 1$, the quantum state preparation is to find an unitary operation $U$ that generates a quantum state
\begin{equation}
    \ket{\psi} = \sum_{i=0}^{2^n-1} \nu_i \ket{i} = U \ket{0}^{\otimes n}
\end{equation}
from the initial state $\ket{0}^{\otimes n}$.
To realized the proposed VQS with parallel Pauli operations, a state preparation circuit can be added for the ancillary qubits.
Consequently, the parallel Pauli operation strategy involves additional computational cost of the state preparation circuit, as a trade-off for improving the scalability of VQS.

The state preparation problem is an active research topic. The computational cost of state preparation has been reduced recently.
The early attempts~\cite{bergholm2005quantum,plesch2011quantum} focused on developing deterministic state preparation algorithms without ancillary qubits.
These attempts resulted in state preparation quantum circuits with $\mathcal{O}(2^n)$ number of CNOT gates.
Recently, the required circuit depth is reduced to $\Theta(n)$ with the exponential number of ancillary qubits~\cite{zhang2022quantum}.
Additional research is needed to further reduce the circuit depth for quantum state preparation.
\section{Conclusion \label{sec:conclusion}}

The proposed QFE framework is a generic and scalable quantum differential equation solver. 
Both stochastic and deterministic differential equations can be solved.
The QFE framework relies on the functional expansion of the solution. 
Therefore, the existence and the convergence of functional expansions in the solution space are the conditions for its applications.
Nevertheless, for most of the differential equations in engineering and scientific problems, spectral methods and interpolation approaches are generally applicable.
The required number of qubits in the QFE framework is directly related to the number of basis functions in the functional expansion.
Only $\mathcal{O}(\log \log (\varepsilon^{-1}))$ number of qubits are required to encode the solution, where the truncation errors reduce exponentially.
Moreover, the scalability of VQS is significantly improved in the framework with the parallel Pauli operation strategy.
In the proposed scalable VQS, the required number of quantum circuits is $\mathcal{O}(M^2)$ regardless of the sparsity of Hamiltonian.
The QFE framework is demonstrated with four example differential equations based on different numerical methods. 
The results show that the QFE framework can solve the differential equations accurately even with two to four qubits.

The solution readout problem of the differential equation solver is not addressed in this paper. In VQS, the solution is stored into a quantum state. However, the solution should be converted to classical information efficiently.
Future work on the quantum state readout to obtain physical quantities of interests is required. 
In addition, reducing the cost of the state preparation circuit in the QFE framework remains as future work.

\appendix

\section{Details on the Chebyshev spectral collocation method.}
\label{appendix:collocation}

The Chebyshev interpolation polynomials in Eq.~(\ref{eq:chebyshev_collocation}) are
\begin{equation}
    S_j(x) = (-1)^{j+1}\frac{(1-x^2)\frac{d T_N(x)}{dx}}{\bar{c}_j N^2 (x-x_j)},
\end{equation}
where 
\begin{equation}
\bar{c}_j = 
    \begin{cases} 
    2 &  (j = 0, N), \\
    1 &  (j = 1,\ldots, N-1),
    \end{cases}
\end{equation} 
and $T_N(x)$ is the $N$-th Chebyshev polynomial of the first kind defined as $T_N(\cos(\theta)) = \cos(N\theta)$.
The Chebyshev differentiation matrix $\mathbf{D}_{N}^{(1)}$ is introduced to linearize the differential operator as
\begin{equation}
    \frac{\partial u(x, t)}{\partial x} \Big|_{x=x_j} \approx \sum_{l=0}^N (\mathbf{D}_N^{(1)})_{jl} u_l(t), \,\, (j=0, \ldots, N).
\end{equation}
The Chebyshev differentiation matrix is obtained by the derivative of Eq.~(\ref{eq:chebyshev_collocation}).
The second differential operator in Eq.~(\ref{eq:heat_eq}) is discretized as the second differentiation matrix
\begin{equation}
(\mathbf{D}_N^{(2)})_{jl}
= \left\{
\begin{array}{ll}
\frac{(-1)^{j+l}}{\bar{c}_l} \frac{x_j^2 + x_j x_l - 2}{(1 - x_j^2)(x_j - x_l)^2}, & j\neq l,\, j\neq 0,\, N \\[10pt]
-\frac{(N^2 - 1)(1 - x_j^2) + 3}{3(1 - x_j^2)^2}, & j = l \neq 0,\, N, \\[10pt]
\frac{2}{3} \frac{(-1)^l}{\bar{c}_l} \frac{(2N^2 + 1)(1 - x_l) - 6}{(1 - x_l)^2}, & j = 0, \, l\neq 0, \\[10pt]
\frac{2}{3} \frac{(-1)^{l+N}}{\bar{c}_l} \frac{(2N^2 + 1)(1 + x_l) - 6}{(1 + x_l)^2}, & j = N, l\neq N  \\[10pt]
\frac{N^4 - 1}{15}, & j = l = 0,\, N.
\end{array}
\right.
\end{equation}
Therefore, Eq.~(\ref{eq:heat_eq}) is converted to
\begin{equation}
    \frac{\partial u_j(t)}{\partial t} = \alpha \sum_{l=0}^N (\mathbf{D}_N^{(2)})_{jl} u_l(t), \,\, j=0, 1, \cdots, N,
\label{eq:appendix_ODEs_heat_equation}
\end{equation}

\section{Details on the Stochastic collocation method}
\label{appendix:scm}

Karhunen-Lo\'{e}ve expansion of $\gamma(x;\omega)$ is
\begin{equation}
    \gamma(x) = \mu(x) + \sum_{i=1}^L \sqrt{\lambda_i} \phi_i(x) \xi_i,
\label{eq:KL_expansion}
\end{equation}
where $\xi_i$'s are independent standard normal Gaussian random variables. $\phi_i(x)$'s and $\lambda_i$'s are eigenfunctions and eigenvalues of covariance function. $\phi_i(x)$'s and $\lambda_i$'s obtained by solving the eigenvalue problem
\begin{equation}
    \int_D C(x,x')\phi_i(x')dx' = \lambda_i \phi_i(x) ,\, x \in D,
\label{eq:eigenproblem}
\end{equation}
where $D$ is the spatial domain. 
In order to solve Eq.~(\ref{eq:eigenproblem}), the eigenfunction $\phi_i(\cdot)$ is chosen as a functional expansion of a basis $\eta_k(\cdot)$ as
\begin{equation}
    \phi_i(x) = \sum_{k=0}^K d_{ik} \eta_k(x) = \sum_{k=0}^K d_{ik} \frac{h_k(x) e^{-x^2 / 2}}{(\sqrt{\pi} 2^k k!)^{1/2}}.
\end{equation}
Here, $h_k(x)$ is a $k$-th Hermite polynomials defined as 
\begin{equation}
    h_{k}(x) = (-1)^{k}e^{x^{2}}{\frac{d^{k}}{dx^{k}}}e^{-x^{2}}.
\end{equation}
The basis $\eta_k(x)$ is weighted Hermite polynomial and normalized to satisfy the orthonormality
\begin{equation}
    \int_{-\infty}^{\infty} \eta_k(x) \eta_m(x) dx = \delta_{km}.
\label{eq:orthonomality}
\end{equation}
Now, Eq.~(\ref{eq:eigenproblem}) is changed into
\begin{equation}
    \int_{-\infty}^{\infty} C(x,x') \sum_{k=1}^K d_{ik} \eta_k(x') dx'  = \lambda_i \sum_{k=1}^K d_{ik} \eta_k(x).
\end{equation}
By using the Eq.~(\ref{eq:orthonomality}),
\begin{align}
\begin{split}
    \sum_{k=1}^K d_{ik} \int_{-\infty}^{\infty} &\int_{-\infty}^{\infty} C(x,x') \eta_k(x') \eta_m(x) dx'dx \nonumber \\
    &= \lambda_i \sum_{k=1}^K d_{ik} \int_{-\infty}^{\infty} \eta_k(x) \eta_m(x) dx
\end{split}
\end{align}
Eq.~(\ref{eq:eigenproblem}) is further organized into
\begin{equation}
    \sum_{k=1}^K d_{ik} \int_{-\infty}^{\infty} \int_{-\infty}^{\infty} C(x,x') \eta_k(x') \eta_m(x) dx' dx = 
    \lambda_i d_{im}.
\label{eq:eigen_problem_organized}
\end{equation}
Let matrix $\mathbf{K}$ as
\begin{equation}
    (\mathbf{K})_{k,m} = \int_{-\infty}^{\infty} \int_{-\infty}^{\infty} C(x,x') \eta_k(x') \eta_m(x) dx' dx.
\end{equation}
Eq.~(\ref{eq:eigen_problem_organized}) is now converted to the eigenvalue problem of $\mathbf{K}$
\begin{equation}
    \mathbf{K}\bm{d}_i = \lambda_i \bm{d}_i.
\end{equation}
Here, $\bm{d}_i$'s are orthonormal eigenvectors, so that $\phi_i$'s are orthonormal eigenfunctions. After solving the eigenproblem of $\mathbf{K}$, the Karhunen-Lo\'{e}ve expansion Eq.~(\ref{eq:KL_expansion}) is obtained.

Stochastic collocation points $\{ \bm{\xi}^{(j)} \}$ are the Gaussian-Hermite quadrature points in Section~\ref{subsec:stochastic_heat}.
For each collocation point $\bm{\xi}^{(j)}$, the deterministic heat equation
\begin{equation}
    \frac{\partial u^{(j)}(x,t)}{\partial t} = \frac{\partial}{\partial x} \left( \gamma^{(j)}(x) \frac{\partial u^{(j)}(x,t)}{\partial x} \right),
\label{eq:deterministic_heat_appendix}
\end{equation}
is solved.
Eq.~(\ref{eq:deterministic_heat_appendix}) can be organized to
\begin{equation}
    \frac{\partial u^{(j)}(x,t)}{\partial t} = \frac{\partial \gamma^{(j)}(x)}{\partial x}  \frac{\partial u^{(j)}(x,t)}{\partial x} + \gamma^{(j)}(x) \frac{\partial^2 u^{(j)}(x,t)}{\partial x^2}.
\end{equation}
For the spatial collocation points $x_{l} = \cos(\pi l/N)$, let $\bm{\gamma} = [\gamma(x_0), \gamma(x_1), \ldots, \gamma(x_N)]^\intercal$, and $\bm{u}(t) = [u(x_0, t), u(x_1, t), \ldots, u(x_N, t)]^\intercal$.
Similar to the example in Section~\ref{subsec:heat}, the Chebyshev differentiation matrix is utilized to convert Eq.~(\ref{eq:deterministic_heat_appendix}) to system of ODEs
\begin{equation}  
    \frac{\partial \bm{u}(t)}{\partial t} = D_N^{(1)} \bm{\gamma} \odot D_N^{(1)} \bm{u}(t) + \bm{\gamma} \odot D_N^{(2)} \bm{u}(t),
\end{equation}
where $\odot$ indicates element-wise multiplication between vectors.
Therefore, the Hamiltonian is
\begin{equation}
    \bm{H} = D_N^{(1)} \bm{\gamma} \odot D_N^{(1)}
    + \bm{\gamma} \odot D_N^{(2)}.
\end{equation}
Mean and variance of $u(x,t)$ are computed from the $u^{(j)}$'s as
\begin{align}
\begin{split}
    \mathbb{E}(u(x,t)) &= \sum_{j} w_j u^{(j)}(x,t) \\
    \mathrm{Var}(u(x,t)) &= \sum_{j} w_j (u^{(j)}(x,t))^2 -\mathbb{E}(u(x,t))^2.
\end{split}
\end{align}


\bibliographystyle{unsrtnat}
\bibliography{apssamp}

\end{document}